\RequirePackage{fix-cm}
\documentclass[twocolumn,epjc3]{svjour3}  
\smartqed  
\RequirePackage{graphicx}
\newcommand{\be}{\begin{equation}}
\newcommand{\ee}{\end{equation}}
\newcommand{\bn}{\begin{eqnarray}}
\newcommand{\en}{\end{eqnarray}}
\newcommand{\bes}{\begin{subequations}}
\newcommand{\ees}{\end{subequations}}

\usepackage{epstopdf}
\usepackage{widetext}
\journalname{Eur. Phys. J. C}
\begin{document}

\title{A complete cosmological scenario from $f(R,T^{\phi})$ gravity theory}


\author{P.H.R.S. Moraes\thanksref{e1,addr1}
        \and
        J.R.L. Santos\thanksref{e2,addr2} 
}

\thankstext{e1}{e-mail: moraes.phrs@gmail.com}
\thankstext{e2}{e-mail: joaorafael@uaf.ufcg.edu.br}


\institute{ITA - Instituto Tecnol\'ogico de Aeron\'autica - Departamento de F\'isica, 12228-900, S\~ao Jos\'e dos Campos, S\~ao Paulo, Brazil \label{addr1}
           \and
          UFCG - Universidade Federal de Campina Grande - Departamento de F\'isica, 58109-970, Campina Grande, Para\'iba, Brazil \label{addr2}
}

\date{Received: date / Accepted: date}

\maketitle

\begin{abstract}
Recent elaborated by T. Harko and collaborators, the $f(R,T)$ theories of gravity contemplate an optimistic alternative to dark energy, for which $R$ and $T$ stand for the Ricci scalar and the trace of the energy-momentum tensor, respectively. Although the literature has shown that the $T$ dependence on the gravitational part of the action - which is due to the consideration of quantum effects - may induce some novel features in the scope of late-time cosmological dynamics, in the radiation-dominated universe, when $T=0$, no contributions seem to rise from such theories. Apparently, $f(R,T)$ contributions to a radiation-dominated universe may rise only from the $f(R,T^{\phi})$ approach, which is nothing but the $f(R,T)$ gravity in the case of a self-interacting scalar field whose trace of the energy-momentum tensor is $T^{\phi}$. We intend, in this article, to show how $f(R,T^{\phi})$ theories of gravity can contribute to the study of the primordial stages of the universe. Our results predict a graceful exit from inflationary stage to a radiation-dominated era. They also predict a late-time cosmic acceleration after a matter-dominated phase, making the $f(R,T^{\phi})$ theories able to describe, in a self-consistent way, all the different stages of the universe dynamics.

\end{abstract}

\section{Introduction}
\label{sec:int}

Plenty of efforts have been made in the theoretical framework with the purpose of explaining the accelerated regime our universe has passed through a fraction of second after the Big-Bang, named ``inflationary era". It is common to describe this phenomenon via scalar fields \cite{guth/1981}-\cite{bezrukov/2014}, although another reputed form to do so comes from the $f(R)$ gravity \cite{nojiri/2007}-\cite{bamba/2014b}. 

Recently, a more general theory of gravity, named $f(R,T)$ gravity, was proposed by T. Harko and collaborators \cite{harko/2011}. {\it A priori} it can also contribute to inflationary era studies through its scalar field approach, named $f(R,T^{\phi})$ gravity \cite{harko/2011}.

Although the late time acceleration of the universe expansion \cite{riess/1998}-\cite{hinshaw/2013} has been broadly investigated in such a theory of gravity \cite{shabani/2013}-\cite{moraes/2015b}, the inflationary era still presents a lack of examination. The same happens for the radiation-dominated era of the universe. 

For the radiation era, the lack of $f(R,T)$ applications is quite predictable for the following reason. The equation of state (EoS) of the universe is $p=\rho/3$ at this stage, with $p$ and $\rho$ representing the pressure and density of the universe, respectively. Such an EoS yields a null trace for the energy-momentum tensor of a perfect fluid. A null trace for the energy-momentum tensor yields the $f(R)$ formalism for the $f(R,T)$ functional forms found in the literature (check, for instance, \cite{harko/2011},\cite{shabani/2013}-\cite{moraes/2015b}). Hence the study of the radiation era of the universe in $f(R,T)$ gravity seems to be quite tricky.

{\it A priori}, it seems reasonable to affirm that the $f(R,T)$ gravity does not contribute to the study of the radiation-dominated universe, since the contribution coming from the trace of the energy-momentum tensor in $f(R,T)$ vanishes at this stage. Specifically, such a shortcoming or incompleteness has attracted attention recently \cite{baffou/2015}-\cite{sun/2015}. 

In \cite{baffou/2015}, the authors argued that the high redshift $f(R,T)$ cosmological solutions tend to recover the standard model of cosmology if the $f(R,T)$ functional form is linear in $R$ precisely because when $z>>1$, the radiation with EoS $p=\rho/3$ dominates the universe dynamics, making the trace of the energy-momentum tensor to vanish. Therefore, no novel contributions would come from the $f(R,T)$ theories of gravitation at the radiation era. 

In \cite{moraes/2014b}, in order to make $f(R,T)$ gravity theory able to contribute also to a radiation-dominated universe, a compactified space-like extra dimension had to be invoked.

In \cite{moraes/2015b}, with the purpose of obtaining well behaved solutions at the radiation stage of the universe, the $f(R,T)$ gravity was generalized, by allowing the speed of light $c$ to vary. Moreover, a scenario alternative to inflation was obtained.

By developing the $f(R,T^{\phi})$ approach, which is submissive to $f(R,T)$ gravity, we intend, here, to make the $f(R,T)$ theories able to contribute also to inflationary and radiation-dominated eras, in a self-consistent way.  Recall that once $T=0$ at the radiation stage, it is natural to expect that $f(R,T)$ theories retrieve $f(R)$ gravity or even general relativity (GR) when $f(R)\sim R$, in such a way one would not expect any new informations derived from $f(R,T)$ gravity. To make the $f(R,T)$ gravity able to describe a radiation universe by itself, without necessarily recovering the $f(R)$ formalism outcomes, will make such a theory able to contribute to all the different stages of the universe dynamics. In order to do so, we will implement the first-order formalism \cite{bazeia/2006,bazeia/2015} to the $f(R,T^{\phi})$ gravity.

We will not omit the recent dynamical features our universe has presented. In other words, a late-time cosmic acceleration, which is corroborated by Type Ia Supernovae observations \cite{riess/1998,perlmutter/1999} and described in $\Lambda$CDM cosmology by the presence of the cosmological constant in standard Einstein's field equations (FEs), shall also be described in our model.

\section{The $f(R,T^{\phi})$ Gravity}

\label{sec:frt}

Early and late time accelerations of the universe expansion can be explained in the framework of scalar field models \cite{guth/1981}-\cite{bezrukov/2014}, \cite{ms/2014}-\cite{das/2006}. The consideration of $f(R,T)$ gravity in the case of a self-interacting scalar field $\phi$ yields the $f(R,T^{\phi})$ gravity \cite{harko/2011}.

Here we will work with the case $f(R,T^{\phi})=-R/4+f(T^{\phi})$, for which $R$ stands for the Ricci scalar, while $T^{\phi}$ is the trace of the energy-momentum tensor of the scalar field, so that

\begin{equation}\label{frt1} 
S^{\phi}=\int d^{4}x\sqrt{-g}\,{\cal L}(\phi,\partial_\mu\,\phi)\,
\end{equation}
is the action of such a field, with $g$ being the determinant of the metric with signature $(+,\,-,\,-,\,-)$ and throughout this article we will assume units such that $4\,\pi\,G=c=1$. Moreover, we are going to deal with a standard Lagrangian density for a real scalar field, whose form is
\be \label{frt2}
{\cal L}=\frac{1}{2}\partial_\mu\phi\partial^{\mu}\phi-V(\phi)\,,
\ee
for which it has been considered a self-interacting potential $V(\phi)$.

Furthermore, since the energy-momentum tensor of the scalar field is given by

\begin{equation}\label{frt3}
T_{\mu\nu}^{\phi}= 2\,\frac{\partial\,{\cal L}}{\partial\,g^{\,\mu\,\nu}}-g_{\,\mu\,\nu}\,{\cal L}\,,
\end{equation}
its trace is 

\begin{equation}\label{frtx1}
T^{\phi}=-\partial_\mu\phi\partial^{\mu}\phi+4V(\phi).
\end{equation} 

The $f(R,T^{\phi})=-R/4+f(T^{\phi})$ gravity model is defined in the following form:

\be\label{eqn:frt4}
S=\int d^{4}x\sqrt{-g}\left[-\frac{R}{4}+f(T^{\phi})+{\cal L}(\phi,\partial_\mu\,\phi)\right],
\ee 
for which we have inserted the gravitational part of the action. By minimizing the above action, we obtain
\be \label{eqn:frt4.5}
G_{\mu\nu}=2[T_{\mu\nu}-g_{\mu\nu}f(T^{\phi})-2f^{\,\prime}(T^{\phi})\,\partial_\mu\phi\,\partial_\nu\phi],
\ee
where $G_{\mu\nu}$ is the Einstein tensor and $f^{\,\prime}$ is the derivative of $f(T^{\phi})$ with respect to $T^{\phi}$. 

For the action $(\ref{eqn:frt4})$, a flat Friedmann-Robertson-Walker universe has the following Friedmann equations:
\be \label{eqn:frt5}
\frac{3}{2}\,H^{\,2}=\left[\frac{1}{2}-2\,f^{\,\prime}(T^{\phi})\right]\,\dot{\phi}^{\,2}-f+V\,,
\ee
\be \label{eqn:frt6}
\dot{H}+\frac{3}{2}\,H^2=-\left[\frac{\dot{\phi}^{\,2}}{2}+f(T^{\phi})\right]+V\,,
\ee
in which $\phi=\phi(t)$, $H=\dot{a}/a$ is the Hubble parameter, $a$ is the scale factor and dots denote derivation with respect to time.

We are also able to derive the equation of motion for such a system, which is given by
\bn\label{eqn:frtxx}
&&
[1-2f^{\,\prime}(T^{\phi})](\ddot{\phi}+3H\dot{\phi}) \\ \nonumber
&&
\hspace{1.5cm}-2\dot{f}^{\,\,\prime}(T^{\phi})\dot{\phi}+[1-4f^{\,\prime}(T^{\phi})]V_\phi=0,
\en
where it was used the notation $V_\phi\equiv dV(\phi)/d\phi$.

\section{First-Order Formalism}
\label{sec:fof}
Let us implement the first-order formalism to the model above, by following the recipe presented in \cite{bazeia/2006,bazeia/2015}. Firstly it is straightforward to observe from $(\ref{eqn:frt5})$ and $(\ref{eqn:frt6})$ that the Hubble parameter obeys the differential equation
\be \label{eqn:fof1}
\dot{H}=-\left(1-2\,f^{\,\prime}\right)\,\dot{\phi}^{\,2}\,.
\ee
Then, the initial ingredient that we may consider in order to establish the first-order formalism is based on the definition
\be\label{eqn:fof1_5}
\dot{\phi} \equiv -W_{\,\phi}\,,
\ee
with $W_{\,\phi}\equiv dW(\phi)/d\phi$ and $W$ is an arbitrary function of the field $\phi$. Such a procedure is commonly used in cosmological scenarios coupled with scalar fields, as we may see in \cite{ms/2014} and references therein. This assumption implies that $(\ref{eqn:fof1})$ is rewritten as
\be\label{eqn:fof2}
\dot{H}=-\left(1-2\,f^{\,\prime}\right)\,W_{\,\phi}^{\,2}\,.
\ee

If we establish that
\be \label{eqn:fof3}
H \equiv h(\phi)+c\,,
\ee
where $c$ is a real constant, then, Eq.$(\ref{eqn:fof1})$ requires the following constraint to the function $h$: 
\be \label{eqn:fof4}
h_\phi=(1-2 f^{\,\prime})\,W_{\,\phi}\,.
\ee
Moreover, the Friedmann equations impose that the potential $V(\phi)$ has to obey the relation
\be \label{eqn:fof5}
V = \frac{3}{2}\,\left(h+c\right)^{\,2}+f\,-\left(\frac{1}{2}-2\,f^{\,\prime}\right)\,W_{\,\phi}^{\,2}.
\ee

\subsection{Examples}

We can apply the first-order formalism by considering a specific form for the function $f(T^{\phi})$. As in \cite{bazeia/2015}, let us deal with $f(T^{\phi})=\alpha\,(T^{\,\phi})^{n}$, with $\alpha$ as a constant and $n$ as an integer. Note that such a functional form for $f(T^{\phi})$ straightforwardly yields the GR case when $\alpha=0$. Moreover, it is the analogous of the form $f(T)=\alpha T^{n}$ in the $f(R,T)$ gravity with no scalar field case, firstly suggested in \cite{harko/2011} for deriving an accelerated cosmological scenario and then applied to a number of other $f(R,T)$ well-behaved cosmological models, such as \cite{singh/2014,moraes/2015}, \cite{moraes/2014b,moraes/2015b}, among many others. Such an assumption yields
\be \label{eqn:fof9}
f(T^{\phi})=\alpha\,\left(-\dot{\phi}^{\,2}+4V\right)^{\,n}=\alpha\,\left(-W_{\,\phi}^{\,2}+4V\right)^{\,n}\,,
\ee
\be \label{eqn:fof10}
f^{\,\prime}=\alpha\,n\left(-\dot{\phi}^{\,2}+4V\right)^{\,n-1}=\alpha\,n\left(-W_{\,\phi}^{\,2}+4V\right)^{\,n-1}.
\ee

In this study, we are going to consider that the function $W$ is given by
\be \label{eqn:fof13}
W(\phi)=a_1\,\sin\,\phi\,,
\ee
with $a_1$ being a real constant. Such a form for $W(\phi)$ is a sine-Gordon type of model, and it has been broadly studied in the literature, specially in investigations related with classical field theory, as one can see in \cite{bazeia/multi} and references therein. Very recently, this model was applied in the context of braneworld scenarios for the $f(R,T)$ gravity  \cite{bazeia/2015}.

Here, for a matter of simplicity, we focus on the case $n=1$, which leads to
\be \label{eqn:fof14}
f^{\,\prime}=\alpha\,.
\ee
Then, by substituting $f^{\,\prime}$ in Eq.$(\ref{eqn:fof4})$, we find
\be \label{eqn:fof15}
h(\phi)=(1-2\,\alpha)\,a_1\,\sin\,\phi\,,
\ee
which means that
\be \label{eqn:fof16}
H=(1-2\,\alpha)\,a_1\,\sin\,\phi+c\,.
\ee
Moreover, it is straightforward to obtain that the potential $V$ is 
\bn
&&
\hspace{-0.3cm}V=\frac{\frac{3}{2}\, \left[(1-2 \alpha) a_1 \sin \,\phi +c\right]^2-\left(\frac{1}{2}-\alpha\right)\, a_1^2 \cos ^2\,\phi }{1-4 \alpha}\,. \\ \nonumber
&&
\en

From Eq.$(\ref{eqn:fof13})$ one can see that the first-order differential equation for this scenario has the form
\be \label{eqn:fof17}
\dot{\phi} = -\,a_1\,\cos\,\phi\,,
\ee
whose analytical solution is
\bn \label{eqn:fof18}
&& 
 \phi(t)=2 \tan ^{-1}\left\{\tanh \left[\frac{1}{2} (b_1-a_1 t)\right]\right\}\,,\\ \nonumber
&&
\en
with $b_1$ being a real constant. The previous solution allows us to rewrite Eq.$(\ref{eqn:fof16})$ as
\be \label{eqn:fof21}
H=(1-2 \alpha)\, a_1 \,\tanh (b_1-a_1 t)+c \,,
\ee
which is presented in details in the upper panel of Fig.\ref{FIG1} below, where we can see that $H$ exhibits a kink-like profile as a direct consequence of the model defined in Eq.$(\ref{eqn:fof13})$.

We can use the last result to determine the expansion parameter (or scale factor) $a(t)$, whose explicit form is given by 
\be \label{eqn:fof22}
a=e^{\,c\,t}\,\left[\mbox{sech} (b_1-a_1 t)\right]^{(1-2 \alpha)}\,.
\ee

\pagebreak

\begin{widetext}
Moreover, the analytical EoS parameter is

\bn \label{eqn:fof23}
&& \nonumber
\\ 
&&
\omega=\frac{\left[12 (\alpha -2) \alpha +7\right]\, a_1^2-3 \left[(1-2 \alpha )^2 a_1^2+c^2\right]\, \cosh [2 (b_1-a_1 t)]+6 \,(2 \alpha -1) \,a_1\, c\, \sinh [2 (b_1-a_1 t)]-3 c^2}{\left[4 (2-3 \alpha ) \alpha -3\right] a_1^2+3 \left[(1-2 \alpha )^2 a_1^2+c^2\right]\, \cosh [2 (b_1-a_1 t)]+6\, (1-2 \alpha )\, a_1\, c \,\sinh [2 (b_1-a_1 t)]+3 c^2}\,, \\ \nonumber
&&
\\ \nonumber
\en

\end{widetext}

\noindent and its features can be observed in the central panel of Fig.\ref{FIG1}. We can verify from the analytical cosmological parameters that small deviations on the values of the constants produce cosmological scenarios which are similar to those presented in Fig.\ref{FIG1}. Therefore, the time evolution of our parameters are not so strongly dependent on the initial conditions of the model.

Another interesting model is 
\be
W(\phi)=a_1\,\left(\frac{\phi^{\,3}}{3}-\phi\,\right),
\ee
whose first-order differential equation is such that
\be
\dot{\phi}=a_1\,\left(1-\phi^{\,2}\right)\,,
\ee
which is satisfied by
\be
\phi=\tanh\,(a_1\,t+b_1)\,,
\ee
where $b_1$ is a real constant. This last first-order differential equation represents the so-called $\phi^4$ type of model, which is also broadly investigated in classical field theory, as one can see in \cite{bazeia/phi4} and references therein. 

In the present case, we are dealing with $f(T^{\,\phi})=\alpha\,T^{\,\phi}$ again. Therefore, we can combine the previous ingredients to determine the following form for the function $h(\phi)$:
\be
h=(1-2\,\alpha)\,a_1\,\phi\,\left(\frac{\phi^{\,2}}{3}-1\right)\,.
\ee
The last equation results in the following relation for the Hubble parameter
\pagebreak
\bn
&& \nonumber
\\ 
&&
H(t)=c-\frac{a_1}{6}(1-2 \alpha )\, \bigg\{3 \sinh (a_1 t+b_1)\\ \nonumber
&&
\hspace{0.9cm}+\sinh \left[3 (a_1 t+b_1)\right]\bigg\}\, \mbox{sech}^3(a_1 t+b_1)\,,
\en
which is presented in details in the upper panel of Fig.\ref{FIG2} below. As in the previous example, it is possible to choose the arbitrary constants in order to obtain a kink-like profile for $H$. 

We can also show that the potential $V$ for this analytical model is
\bn
&&
V=\frac{1}{1-4 \alpha}\,\bigg\{\frac{1}{6}\, \left[(1-2 \alpha)\, a_1\, \phi  \left(\phi ^2-3\right) +3 \,c\right]^2\\ \nonumber
&&
\hspace{0.5cm}-\left(\frac{1}{2}-\alpha\right)\, a_1^2 \left(\phi ^2-1\right)^2\bigg\}\,.
\en

Withal, the expansion parameter for such a case is given by the following

\bn
&&
a=\exp\,\left[c\,t+\frac{1}{6}\,(1-2\, \alpha)\, \mbox{sech}^2(a_1 t+b_1)\right] \\ \nonumber
&&
\hspace{0.3cm}\times\,\left[\cosh (a_1 t+b_1)\right]^{\,\frac{2}{3}(2\alpha-1)}\,,
\en
while the EoS parameter reads

\begin{widetext}
\bn
&& \nonumber
 \omega=-1+192 (1-4 \alpha) a_1^2 \cosh ^2(a_1 t+b_1)\,\bigg\{ -8 \left[4 \alpha  (5 \alpha -2)+5\right] a_1^2+3 \left[4 (4 (\alpha -3) \alpha +1) a_1^2+45 c^2\right]\, \cosh [2 (a_1 t+b_1)] \\ \nonumber
&&
\hspace{0.3cm}+\,\,6 \,\left[4 (1-2 \alpha )^2\, a_1^2+9\, c^2\right]\, \cosh [4\, (a_1 t+b_1)]+16 \,\alpha ^2 a_1^2 \,\cosh [6\, (a_1 t+b_1)]-16\, \alpha \, a_1^2\, \cosh [6\, (a_1 t+b_1)] \\ \nonumber
&&
\hspace{0.3cm}+\,\,4\, a_1^2 \,\cosh [6\, (a_1 t+b_1)]+9\, c^2 \,\cosh [6\, (a_1 t+b_1)]+216\, \alpha\,a_1\, c\, \sinh [2\, (a_1 t+b_1)]+144 \,\alpha \, a_1 \,c\, \sinh [4\, (a_1 t+b_1)]\\ 
&&
\hspace{0.3cm}+\,\,12\, (2 \alpha -1)\, a_1\, c\, \sinh [6\, (a_1 t+b_1)]-108\, a_1\, c\, \sinh [2\, (a_1 t+b_1)]-72\, a_1 \,c\, \sinh [4 \,(a_1 t+b_1)]+90\, c^2\bigg\}^{\,-1}\,.
\en

\end{widetext}

\noindent The latter is plotted in the central panel of Fig.\ref{FIG2}. Again, if we take small deviations of these parameters we determine similar cosmological scenarios.

\begin{figure}[ht!]
\vspace{0.3cm}
\centering
\includegraphics[width=0.8\columnwidth]{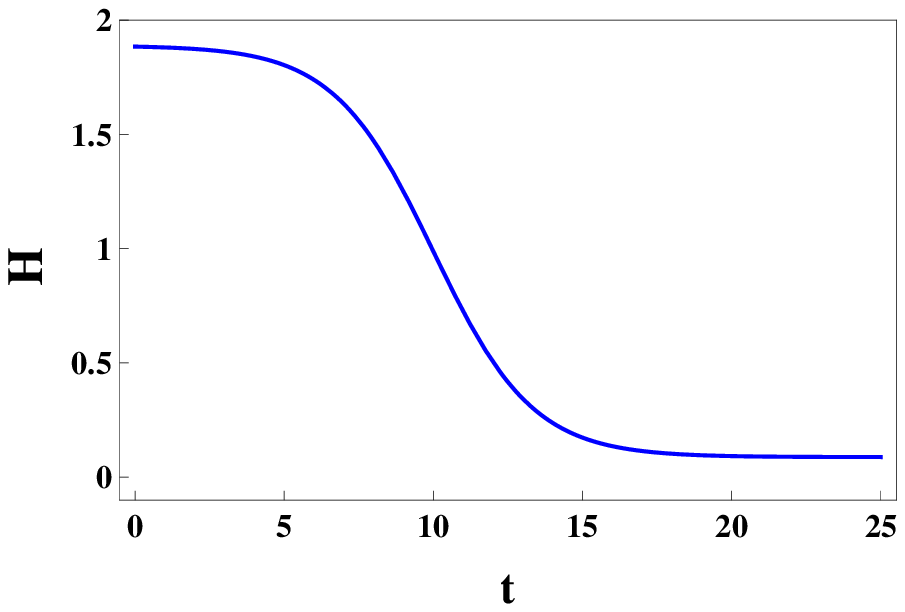}
\vspace{0.2cm}
\includegraphics[width=0.8\columnwidth]{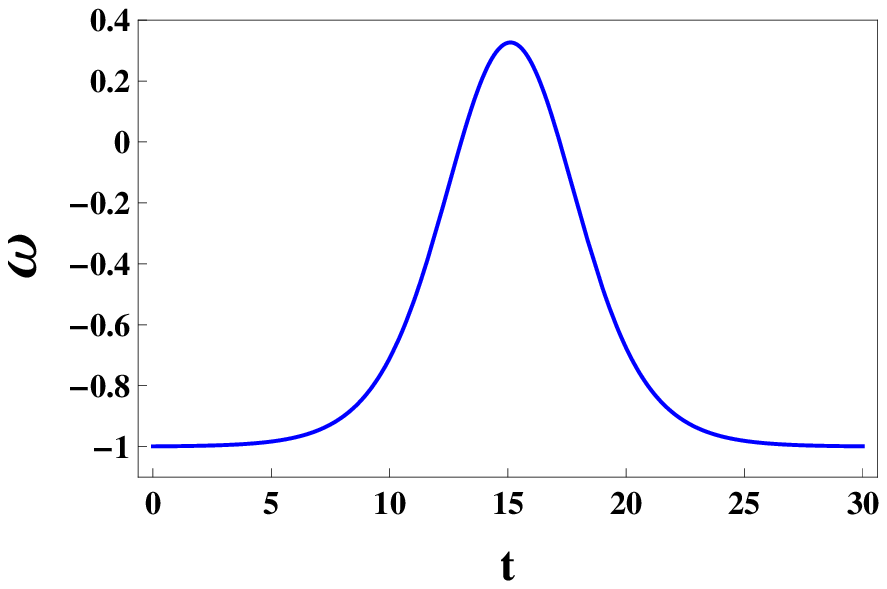}
\vspace{0.3cm}
\includegraphics[width=0.8\columnwidth]{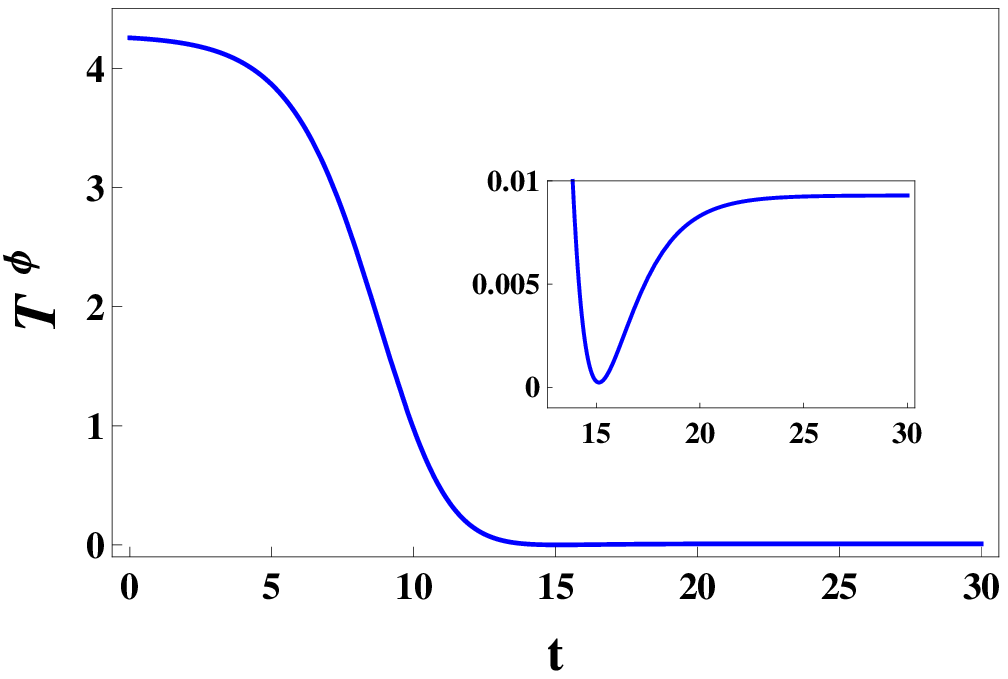}
\vspace{0.2cm}
\caption{The upper panel shows the evolution of $H(t)$, the central panel shows $\omega$, and the lower panel shows $T^{\,\phi}$, for $\alpha =-1$,  $a_1 = 0.3 $, $b_1 = 3$, and $ c = 0.988$, in the $W(\phi)=a_1\,\sin\,\phi\,$ model.}
\label{FIG1} 
\end{figure}

\begin{figure}[hb!]
\vspace{0.3cm}
\centering
\includegraphics[width=0.8\columnwidth]{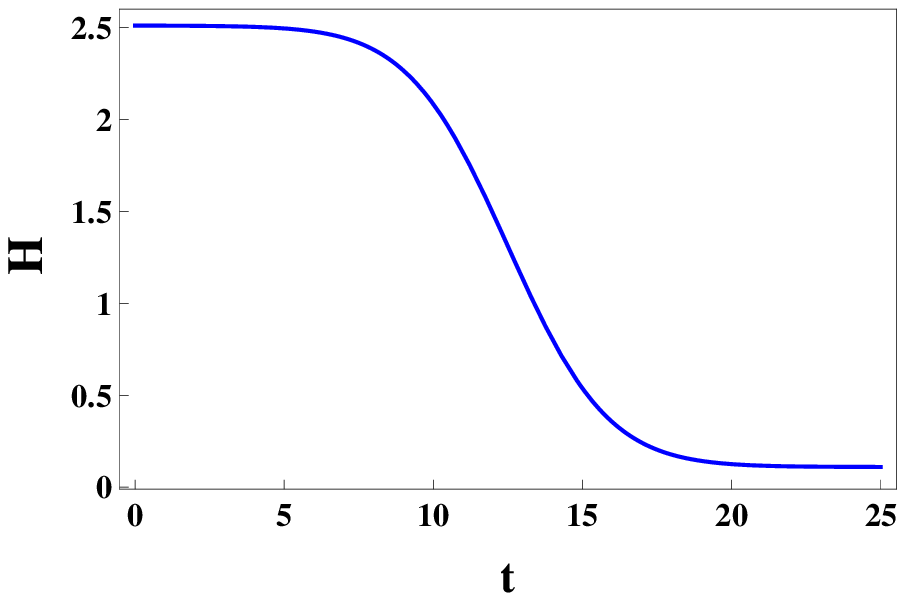}
\vspace{0.2cm}
\includegraphics[width=0.8\columnwidth]{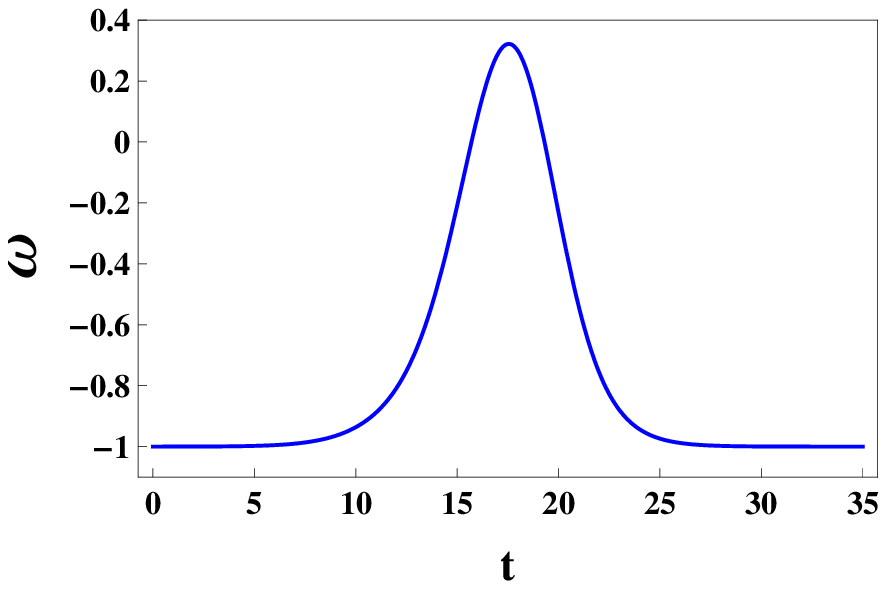}
\vspace{0.3cm}
\includegraphics[width=0.8\columnwidth]{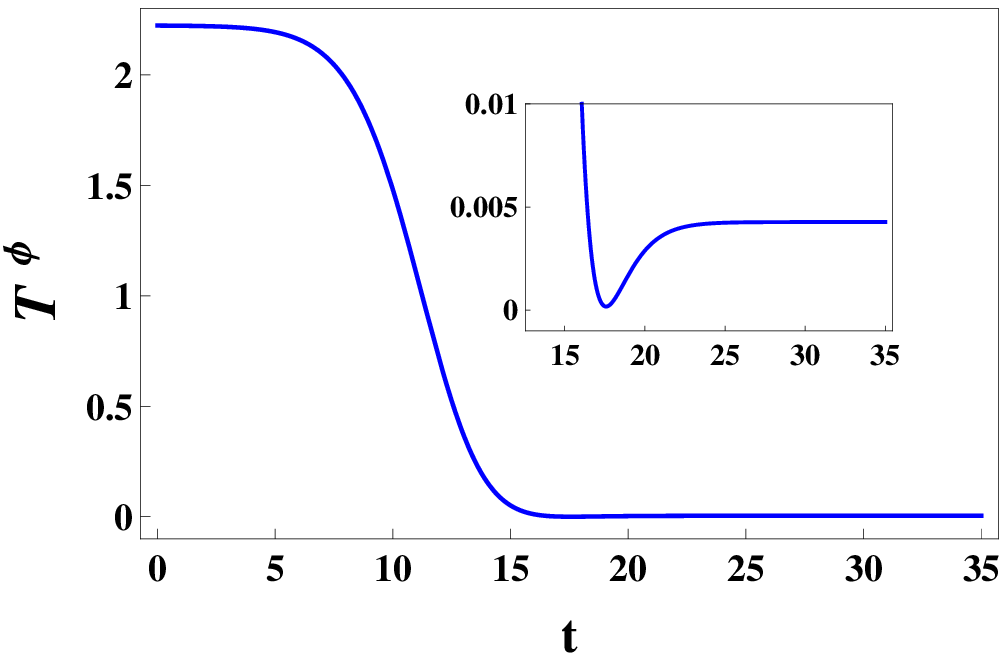}
\vspace{0.2cm}
\caption{The upper panel shows the evolution of $H(t)$, the central panel shows $\omega$, and the lower panel shows $T^{\,\phi}$, for $\alpha =-4$,  $a_1 = 0.2 $, $b_1 = -2.5$, and $ c = 1.31$, in the $W(\phi)=a_1\,(\phi^{\,3}/3-\phi\,)$ model.}
\label{FIG2} 
\end{figure}	

\section{Cosmological Interpretations}

\label{sec:ci}

In Section \ref{sec:fof}, we were able to depict the time evolution of $H$ and $\omega$ derived from the first-order formalism applied to the $f(R,T^{\phi})$ theory. Remind that our aim in this work is to make $f(R,T)$ gravity able to induce a complete cosmological scenario, which includes the primordial eras of the universe, since the theory seems to present some shortcomings at these stages. Therefore, it is worth checking if the cosmological parameters derived previously indeed resemble, for small values of time, what is expected for the primordial universe dynamics. 

In the previous section we have also plotted the evolution of the trace of the energy-momentum tensor of the scalar field, which shall be interpreted below. 

The inflationary model states that early in the history of the universe, its expansion has accelerated. From standard Friedmann equations, this happens if the universe is dominated by a component with EoS $\omega<-1/3$ \cite{ryden/2003,dodelson/2003}. According to standard cosmology, during inflation, the energy density of the universe was dominated by a constant, say, $\Lambda_\iota$, in such a way the Friedmann equations read $H^{2}_\iota\sim\Lambda_\iota$. The Hubble parameter was, then, constant during inflationary era. Indeed, one may write, for inflation, $a_\iota\sim e^{H_\iota t}$ \cite{ryden/2003,dodelson/2003}.

From the upper panels of Figs.\ref{FIG1}-\ref{FIG2}, the Hubble parameter values predicted from our model indeed remain approximately constant during a small period of time after the Big-Bang, in agreement with inflationary scenario. In fact, $\dot{H}\sim -0.003$ and $\sim -0.0003$ at $t = 0.01$ for Figs.\ref{FIG1} and \ref{FIG2}, respectively.

After inflation, $H$ must evolve as $\propto t_H^{-1}$, with $t_H$ being the Hubble time \cite{ryden/2003,dodelson/2003}, in accordance with Figs.\ref{FIG1}-\ref{FIG2}.

Still in the upper panels of Figs.\ref{FIG1}-\ref{FIG2}, one should note that for high values of time, $H$ behaves once again as a constant. From standard cosmology, it is well known that for high values of time, $H$ is proportional to $\rho_\Lambda$ (the density of the cosmological constant), which is, indeed, a constant. Therefore, the recent cosmic acceleration our universe is undergoing can be described by a scale factor which evolves as $e^{H_0t}$, with the constant behaviour of $H$ being predicted in Figs.\ref{FIG1}-\ref{FIG2} for high values of time.

In the central panels of Figs.\ref{FIG1}-\ref{FIG2}, we have depicted the evolution of the EoS parameter $\omega$ with time. By analysing it, one can see that for small values of time, $\omega\sim-1$. This is in agreement with some constraints that have been put to inflationary EoS recently \cite{bamba/2013}-\cite{aldrovani/2008}.

After representing inflation, $\omega$ smoothly evolves to $\sim1/3$. This is the maximum value the EoS should assume during the universe evolution. Moreover, $1/3$ is precisely the value of the EoS which describes the radiation-dominated era of the universe \cite{ryden/2003,dodelson/2003}, i.e., the model predicts a smooth transition from inflationary stage to radiation-dominated era.

As time passes by, the central panels of Figs.\ref{FIG1}-\ref{FIG2} show that $\omega\rightarrow-1$. According to recent observations of anisotropies in the temperature of cosmic microwave background radiation \cite{hinshaw/2013}, this is the present value of the universe EoS, and is responsible for the recent cosmic acceleration.

The fact that nowadays the universe is passing through its second phase of accelerated expansion justifies such an evolution for $\omega$. Since an accelerated universe is described by a negative EoS, it is expected that during the whole universe evolution, the value of $\omega$, coming from $\sim-1$, which is related to the inflationary stage, first increases, tending to a decelerating phase, then decreases, returning to an accelerated phase.

Furthermore, the lower panels of Figs.\ref{FIG1}-\ref{FIG2} show the time evolution of the trace of the energy-momentum tensor of the scalar field for the models.

Firstly, one can note that the values assumed by such a quantity are always $\geq0$, as they, indeed, should be.

The high values of density and temperature which are known to characterize the Big-Bang, together with the inflationary EoS mentioned above, make us to expect to have the maximum values of the trace of the energy-momentum tensor of the scalar field during inflation, as corroborated by Figs.\ref{FIG1}-\ref{FIG2}.

At the radiation era, $T^{\phi}$ must vanish. We zoomed in on Figs.\ref{FIG1}-\ref{FIG2} lower panels in order to explicit such a predicted feature. Note that after a period assuming its maximum values, the trace of the energy-momentum tensor $T^{\phi}\rightarrow0$ at the same time scale $\omega\rightarrow1/3$ in Figs.\ref{FIG1}-\ref{FIG2} central panels. Such a property reinforces the good behaviour of our solutions, specifically at the primordial stages of the universe evolution.

At matter-dominated era, $T^{\phi}$ must be $\neq0$. The same happens for the second period of acceleration. Indeed, the plots of $T^{\phi}$ show that after the radiation era, $T^{\phi}$ increases its values. Still, such values are small, due to the low density of the universe at these stages, specially when compared to the $\rho$ values at inflation.

Remarkably, $T^{\phi}$ remains constant for high values of time. Since $T^{\phi}\propto\rho$, this characterizes the late-time universe dynamics to be dominated by a constant. Recall that in standard cosmology, for high values of time, the cosmological constant $\Lambda$ dominates the universe dynamics. As argued above, $\rho_{\Lambda}$ remains, indeed, constant as time passes by. Such an important feature of standard cosmology is predicted from the $f(R,T^{\phi})$ approach here presented.

\section{Discussion and perspectives}
\label{sec:dis}

In this article we established a first-order formalism to determine two analytical models related to the $f(R,T^{\,\phi})$ theory. Such a formalism was constructed from the definition observed in Eq.(\ref{eqn:fof1_5}). 

The mentioned expression together with Eq.(\ref{eqn:frt5}) and (\ref{eqn:fof3}) gave us a constraint for the potential $V$, which can be viewed in (\ref{eqn:fof5}). Moreover, the arbitrary function  $h(\phi)$ was introduced and its form is directly related to $f^{\,\prime}$.

The first analytical model was generated by considering $W=a_1\,\sin\,\phi$ and the specific form for $f(T^{\,\phi})$ showed in $(\ref{eqn:fof9})$. Such definitions leaded us to determine $h(\phi)$, besides the potential $V$, the Hubble parameter $H$, as well as other cosmological quantities like the expansion and EoS parameters. In the second case, we adopted $W=a_1\,(\phi^{\,3}/3-\phi)\,$, and one more time we determined the function $h$ and the cosmological parameters for the model.

We can directly see that the field $\phi(t)$ satisfies the equation of motion (\ref{eqn:frtxx}) for both models. Furthermore, this first-order formalism may help us to obtain other one-field analytical scenarios and also may lead us to hybrid analytical models, where it is considered the coupling between an $f(R,T^{\phi,\chi})$ function with a two scalar field Lagrangian. Such a scenario shall be derived and presented in near future. 

Another further work may rise from an alternative form for $f(R,T^{\phi})$, such as $f(R,T^{\phi})=g(R)h(T^{\phi})$, with $g(R)$ and $h(T^{\phi})$ being functions of $R$ and $T^{\phi}$, respectively. Such a functional form suggests a high coupling between matter and geometry, and therefore may imply some novelties in gravitational and cosmological perspectives. We can conjecture that by coupling such kind of theories with a scalar field lagrangian may lead us to a new set of Friedmann equations and to different equations of motion. Once we have these ingredients we may try to apply the first-order formalism and then compute the constraints that the superpotential will obey. In fact, the analogous of that model for the $f(R,T)$ gravity with no scalar field case was investigated in \cite{shabani/2013}.

We have obtained from the first-order formalism applied to the $f(R,T^{\phi})$ gravity, two models able to describe all the different dynamical stages of the universe, from inflation, to radiation, matter and dark-energy dominated eras. 

For the first time, an $f(R,T)$ cosmological model was able to describe all the dynamical stages of the universe continuously, including the primordial ones. There was no need of invoking different EoS or functional forms for $f(T^{\phi})$ in different stages in order to be able to account for all the transient eras of the universe dynamics, as made in \cite{sun/2015}, \cite{reddy/2013} and \cite{ram/2013}, for instance. In fact, there was no need of invoking any EoS at all. Such a quantity was obtained simply by dividing the solutions $p$ by $\rho$.

It is remarkable the fact that the inflationary scenarios of our models have smoothly decayed to a radiation-dominated era. This phenomenon is sometimes called ``graceful exit" and it is not trivial to be obtained (check \cite{lima/2013,perico/2013} for the graceful exit achievement through decaying vacuum models). In fact, an ideal inflationary scenario must naturally decay to a radiation-dominated era \cite{lima/2015,basilakos/2013}, as in the present case.

Also remarkable is the fact that the models are well-behaved at radiation stage. The match between the time scales for $\omega=1/3$ e $T^{\phi}=0$ in Figs.\ref{FIG1}-\ref{FIG2} reinforces that. 

Not just the present models were able to describe a smooth transition between inflationary and radiation-dominated eras, but they were also able to predict the matter dominated phase, with $\omega=0$, and the recent cosmic acceleration the universe is undergoing \cite{riess/1998}-\cite{hinshaw/2013}, from an EoS parameter whose late-time values agree with WMAP cosmic microwave background observations \cite{hinshaw/2013}.

Therefore, departing from a cosmological scenario derived from GR, we have obtained an EoS which varies with time, allowing, through its evolution, the universe dynamics to be dominated by different components in a natural form. Moreover, in order to obtain a recent cosmic acceleration, we did not have to insert a cosmological constant in the model FEs, in this way, evading the cosmological constant problem \cite{weinberg/1989,peebles/2003,padmanabhan/2003}.

The $f(R,T^{\phi})$ cosmological scenario derived in this work is somehow similar to a decaying vacuum $\Lambda(t)$ model, such as those presented in \cite{lima/2013,perico/2013}, for instance. Firstly, it should be stressed that recently in \cite{sun/2015}, the $f(R,T)$ theories have been interpreted as a generalization of the Brans-Dicke model \cite{brans/1961}-\cite{tripathy/2015}. As argued by the authors in \cite{sun/2015}, the extra terms in the $f(R,T)$ FEs may play the role of additional terms of effective gravitational and cosmological ``constants" $G_{eff}$ and $\Lambda_{eff}$. Those extra terms in $\Lambda_{eff}$ are the responsible for a vacuum decay in $\Lambda(t)$-like models. 

Second, in both decaying vacuum and $f(R,T)$ models, the covariant derivative of the energy-momentum tensor $\nabla^{\mu}T_{\mu\nu}$ does not vanish. In decaying vacuum models, this requires necessarily some energy exchange between matter and vacuum, through vacuum decay into matter or vice versa. 

The vacuum-decay models presented in \cite{lima/2013,perico/2013}, for instance, are capable of describing the complete history of the universe in an effective unified framework, in the same way the present $f(R,T^{\phi})$ model does. In those models, the vacuum always decays in the dominant dynamical component of each stage. However, in order to be able to describe the various phases of the universe dynamics, the functional form taken for $\Lambda(t)$ is proposed on phenomenological grounds, by extrapolating backwards in time the present available cosmological data. On the other hand, the $f(R,T^{\phi})$ scenario is free of ansatz or phenomenology.

Note also that the $\Lambda(t)$ model presented in \cite{lima/2013} was mimic through a scalar field model (check Section 4 of such a reference). This corroborates the argument that it might have some fundamental analogy between scalar field and decaying vacuum models. 

In this article, we have also formally brought to the scientific community attention the shortcoming surrounding the $f(R,T)$ theories of gravity in the regime $T=0$. It is, indeed, straightforward to realize that in such a regime of $f(R,T)$ gravity one automatically recovers the $f(R)$ theories, or GR when $f(R)$ is linear in $R$ \cite{baffou/2015}. We have shown, as an unprecedented outcome, that it is possible to describe radiation in $f(R,T)$ gravity not necessarily by the recovering of GR, but through extra dynamical terms coming from a scalar field in the $f(R,T^{\phi})$ approach.

If one focus specifically on the radiation era predicted by the $f(R,T^{\phi})$ model here presented, one is able to obtain a scale factor which evolves as the one of standard cosmology purely from the $f(T^{\phi})$ contribution, which is a complementary novel feature of the $f(R,T)$ gravity that we will show below. 

By taking $T^{\phi}=0$ in (\ref{frtx1}) yields $\dot{\phi}^{2}=4V(\phi)$. Since in this case $f(T^{\phi})=\alpha (T^{\phi})^{n}=0$, one obtains, from (\ref{eqn:fof1}), 

\begin{equation}\label{c1}
\dot{H}=-4V(\phi).
\end{equation}
From Eqs.(\ref{eqn:fof1_5}), (\ref{eqn:fof5}) and (\ref{c1}), one can write $\dot{H}=-2H^{2}$, which integrated for the scale factor $a$ yields $a\sim t^{1/2}$. From standard cosmology \cite{ryden/2003,dodelson/2003}, this is exactly the time proportionality of the scale factor of a radiation-dominated universe. Therefore the formalism presented in this article surpasses the $f(R,T=0)$ issue since it makes $f(R,T)$ gravity able to describe a radiation-dominated universe purely from its dependence on $T^{\phi}$.

Besides some questions exposed in the Introduction, very recently, the drawback in $f(R,T)$ theory was raised in \cite{sun/2015}, in which the authors have searched for different forms for $f(T)$ in different stages of the universe. In the radiation-dominated stage, they found that $f(T)$ should be a non-null constant. As the authors have considered the $f(R,T)$ theories as a generalization of Brans-Dicke gravity, the $f(T)\sim cte$ case leads to $G_{eff}$ and $\Lambda_{eff}$  to be, indeed, constants, as it may be seen in \cite{sun/2015}. Therefore no contributions to a $T=0$ universe would come from the $f(R,T)$ formalism on this approach, departing from the results of the present work.  

One could argue that the $f(R)$ formalism retrieval in $f(R,T)$ theories when $T=0$ is not a shortcoming, being just an intrinsic property of such theories. However, below we bring to the reader's attention some fundamental areas of knowledge which will have to be neglected if the $f(R,T)$ theories, indeed, simply recovers the $f(R)$ formalism when $T=0$. 

Many contributions to Physics, Astrophysics and Cosmology will come from the imminent detection of gravitational waves (GWs) \cite{aasi/2015,acernese/2015}. Their detection will clarify some persistent and important issues, such as the absolute ground state of matter \cite{moraes/2014}, and also contemplates a powerful tool in estimating parameters of compact binary systems \cite{berry/2015}, constraining the equation of state of neutron stars \cite{takami/2014} and brane cosmological parameters values in braneworld models \cite{mm/2014}, and distinguishing GR from alternative theories of gravity \cite{alves/2009}.

The study of the propagation of GWs is made once they have been generated \cite{buonanno/2007}. The energy-momentum tensor $T_{\mu\nu}$ is then set to zero in order to obtain the wave equation for vacuum \cite{wang/2009}-\cite{dzhunushaliev/2014}. In the same way the EoS $\omega=1/3$ yields a null trace for the energy-momentum tensor, this will happen for vacuum, with $T_{\mu\nu}=0$. Therefore it is expected that the study of GWs, as the derivation of their primordial spectrum, for instance, in the $f(R,T)$ formalism yields the same predictions as those obtained via $f(R)$ gravity \cite{corda/2010}. 

A powerful tool to study the properties of GWs was developed in \cite{eardley/1973}. It consists in characterizing the polarization states of GWs in a given theory\footnote{In general relativity, there are two polarization states of gravitational waves, known as plus and cross polarizations.}. In order to do so, one evaluates the Newman-Penrose (NP) quantities \cite{newman/1962,newman/1962b}. In \cite{alves/2009}, from the NP quantities calculation, the authors have found extra polarization states for the $f(R)$ gravity. Those extra states may be corroborated by experiment once the advanced GW detectors, such as LIGO \cite{aasi/2015} and VIRGO \cite{acernese/2015}, start they next run. Since the NP formalism is applied to the vacuum FEs of a given theory, such an application to the $f(R,T)$ theory is expected to yield the same extra polarization states as those of the $f(R)$ formalism \cite{alves/2009}.

The Kerr metric describes the geometry of an empty space $(T_{\mu\nu}=0)$ around a rotating uncharged black hole (BH) with axial symmetry. Kerr BHs studies have been constantly seen in the literature (check, for instance, \cite{herdeiro/2014}-\cite{bhadra/2012}). Since BHs are characterized by strong gravitational fields in quantum length scales, it is expected that their physics bring up some new insights about quantum theories of gravity \cite{rovelli/1996,ashtekar/1998}. Once again, one would not obtain new contributions to the study of the space-time around a Kerr BH from the $f(R,T)$ formalism, in such a way the $f(R)$ theories outcomes should be expected \cite{myung/2011}-\cite{cembranos/2011}.

Hence, one can see that the lack of contributions from $f(R,T)$ gravity will appear not only in a radiation-dominated universe, but also in studies made in the vacuum regime, since $T_{\mu\nu}=0\rightarrow T=0$. In other words, in the absence of the $f(R,T^{\phi})$ approach presented here, one would expect the $f(R)$ formalism features retrieval in the study, not only of a radiation-dominated universe, but also in those of GWs and space-times surrounding BHs. 

For instance, in the search for new polarization states of GWs mentioned above, one works with the vacuum FEs of a theory. It is expected that if one considers the vacuum on the rhs of standard $f(R,T)$ FEs (check Eq.(11) of \cite{harko/2011}), one gets, after the NP formalism application, the same polarization states of the $f(R)$ theory \cite{alves/2009}. On the other hand, by taking $T_{\mu\nu}=0$ (and consequently $T=0$) in the present model FEs (\ref{eqn:frt4.5})\footnote{Remind that we have assumed $f(R,T)$ is linear in $R$, so that the lhs of Eq.(\ref{eqn:frt4.5}) is purely the Einstein tensor, as in general relativity.}, one expects the third term inside the brackets to survive in the vacuum regime. This term should naturally yield some departure from both GR and $f(R)$ gravity in the study of GWs polarization states. Moreover, it could be responsible for novelties in other astrophysical areas submissive to the GW subject, such as the characterization of the GW spectrum \cite{buonanno/2007}, the evolution of cosmological perturbations \cite{mukhanov/1992} and the GW background from cosmological compact binaries \cite{farmer/2003}.  

Furthermore, another vacuum studies in $f(R,T)$ gravity will be allowed to be investigated, such as the Tolman-Oppenheimer-Volkoff-de Sitter equation \cite{makino/2015} and the vacuum polarization by topological defects \cite{bezerra_de_mello/2009}, among many others. Those points above justify the search for evading the $f(R,T=0)$ issue, which was presented in this article.

The contributions from the present work are able to motivate new branches in the study of the radiation universe and vacuum in $f(R,T)$ gravity. The present formalism makes possible to make the most of the $f(R,T)$ theories. Although the $f(R)$ formalism retrieval in the $T=0$ regime would not necessarily be a critical feature of the $f(R,T)$ theories, it would make us overlook plenty of applications of those theories, which may generate some remarkable outcomes. Some of those, as possible extra polarization states of GWs, may be inferred by experiment soon.\\

\begin{acknowledgements}
PHRSM would like to thank S\~ao Paulo Research Foundation (FAPESP), grant 2015/08476-0, for financial support. The authors are grateful to Drs. M.E.S. Alves and J.C.N. de Ara\'ujo for their suggestions in what concerns the physical interpretation of the present model and to the anonymous referee for his/her suggestions which certainly enriched the physical and cosmological content of the paper.  
\end{acknowledgements}



\end{document}